\documentstyle[aps,prb,epsf]{revtex}
\draft
\begin{document}

\title{Density-Matrix Renormalization-Group Analysis of Quantum 
Critical Points: \\
I. Quantum Spin Chains}

\author{Shan-Wen Tsai and J. B. Marston}

\address{Department of Physics, Brown University, Providence, RI 02912-1843}

\date{\today}  

\maketitle

\begin{abstract}
We present a simple method, combining the density-matrix 
renormalization-group (DMRG) algorithm with finite-size scaling, 
which permits 
the study of critical behavior in quantum spin chains.  Spin moments and 
dimerization are induced by boundary conditions at the chain ends and these 
exhibit power-law decay at critical points.  Results are presented for the
spin-$1/2$ Heisenberg antiferromagnet; an analytic calculation shows that
logarithmic corrections to scaling can sometimes 
be avoided.  We also examine the 
spin-$1$ chain at the critical point separating the Haldane gap and 
dimerized phases.  Exponents for the dimer-dimer and the spin-spin 
correlation functions are consistent with results obtained from bosonization.
\end{abstract}

\pacs{PACS numbers: 75.10.Jm, 05.70.Jk, 75.40.Mg}

\section{Introduction}
\label{sec:Intro}
Quantum critical points are characterized by fluctuations over all length 
and time scales and by the appearance of power law scaling.  In this paper
we present a simple but powerful numerical 
method to access quantum critical points 
in one-dimensional systems.  The method combines the 
density-matrix renormalization-group (DMRG) algorithm and 
finite-size scaling ideas.  We illustrate the method by applying it to 
several well-understood quantum spin chains.  In a second paper to follow
we apply the method to new classes of supersymmetric spin chains which 
describe various disordered electron systems\cite{part2}.

The development of the density-matrix renormalization-group (DMRG) 
algorithm by White\cite{White} represented an 
important improvement over previous numerical methods for the study of low 
dimensional lattice models.  It has been applied to a wide variety 
of systems\cite{proceedings}. 
The DMRG approach was first used to study the ground state properties and 
low-energy excitations of one-dimensional chains.  It has been extensively 
applied to the study of various spin chains.  Low-lying excited states of the 
spin-$1$\cite{White2,Sorensen1,Sorensen2} and spin-$1/2$\cite{Ng} Heisenberg 
antiferromagnets have been calculated.  Likewise, spin-$1$ chains 
with quadratic and biquadratic interactions\cite{Bursill,Fath}, a spin-$2$ 
antiferromagnetic chain\cite{Schollwock,Wang1}, spin-$1/2$ and spin-$1$ 
chains with dimerization and/or frustration 
(next-nearest-neighbor coupling)\cite{Kato,Bursill1,Chitra,Kolezhuk,Watanabe}, 
and frustrated spin-$3/2$ and spin-$2$ chains\cite{Roth} have all been studied.
Edge excitations\cite{Sorensen2,Schollwock,Qin,Polizzi} at the ends of 
finite spin chains and the effects of perturbations such as 
a weak magnetic field coupled to a few sites\cite{Legeza} have been considered. 
Randomness in the form of random transverse magnetic field in a spin-$1/2$ XY 
model\cite{Juoza}, random exchange couplings\cite{Hida1}, and random 
modulation patterns of the exchange\cite{Schonfeld,Hida2}, has been 
examined.  Finally, alternating spin magnitudes\cite{Pati}, the presence of a 
constant\cite{Venuti} or a staggered\cite{Lou} magnetic field in a spin-$1$ 
chain, bond doping\cite{Wang2}, the effects of a local 
impurity\cite{Sorensen3}, and interactions with quantum 
phonons\cite{Caron,Bursill2} have also been considered. 

Most of the above work involves systems in which the 
first excited state is separated from the ground state by a non-zero energy
gap as the DMRG works best for gapped systems.  First attempts to extract 
critical behavior of gapless systems used the DMRG to generate 
renormalization transformations of the coupling constants in the 
Hamiltonian\cite{Drzewinski,Bursill3}.  Hallberg et al.\cite{Hallberg} 
studied the critical behavior of $S=1/2$ and $S=3/2$ quantum spin chains with 
periodic boundary conditions through extensive calculations of ground state 
correlation functions at different separations and different chain sizes 
$L$.  Spin correlation functions in an open chain have also been calculated 
and compared with results calculated from low-energy field theory, showing 
that estimates of the amplitudes can also be obtained\cite{Hikihara}.  The 
approach described in this paper was applied to the spin-$1/2$ Heisenberg 
chain and a non-Hermitian supersymmetric (SUSY) spin 
chain\cite{Kondev}.  More recently, 
critical behavior of classical one-dimensional reaction-diffusion 
models\cite{Carlon1} and the two-dimensional Potts model\cite{Carlon2} has 
been studied using the finite-size DMRG algorithm.  Bulk and surface 
exponents of the Potts and Ising model have been obtained by using the DMRG 
to calculate correlation functions at different separations and collapsing 
curves obtained at different system sizes\cite{Kaulke}.  The SUSY chain 
describing the spin quantum Hall effect (SQHE) plateau transition was also 
examined in some detail.  Critical exponents were extracted\cite{Senthil} 
and compared to exact predictions\cite{Gruzberg}.  Thermodynamic properties 
of other two-dimensional classical critical systems have also been studied 
by the DMRG method\cite{Nishino,Carlon3,Honda}.  Finally, Andersson et al. 
investigated the convergence of the DMRG in the thermodynamic limit for a 
gapless system of non-interacting fermions\cite{Andersson}.

The method described in this paper combines the DMRG algorithm 
with finite-size scaling analysis, and yields accurate critical exponents. 
The main advantage of the method is its simplicity.  Only the 
calculation of ground state correlations near the middle of chains with 
open boundary conditions are required.  The relatively simple ``infinite-size'' 
DMRG algorithm\cite{White} is particularly accurate for this job.  
In Sec. \ref{sec:method} we describe the method.  The tight-binding model 
can be solved exactly and in Sec. \ref{sec:tightbinding} we use it to 
illustrate our scaling analysis.  DMRG results are presented in 
Sec. \ref{sec:s=half} for the anisotropic $S=1/2$ Heisenberg 
antiferromagnet and several critical exponents are obtained.  An analytical 
calculation shows that multiplicative logarithmic corrections -- which 
complicate the extraction of accurate critical exponents -- may be avoided
in some instances.  In Sec. \ref{sec:s=1}, 
the $S=1$ antiferromagnetic spin chain is studied, focusing on the critical 
point that separates the Haldane and the dimerized phases.  We conclude 
with a summary in Sec. \ref{sec:conclusion}.

\section{The DMRG / Finite-Size Scaling Approach}
\label{sec:method}
We first describe how critical exponents may be obtained from a finite-size 
scaling analysis of chains with open or fixed boundary conditions.  These
boundary conditions are the simplest to implement in DMRG calculations. 
In the next subsection the DMRG algorithm itself is briefly described.
 
\subsection{Finite-Size Scaling}
\label{subsec:finite}
To illustrate the sorts of power-law scaling we wish to examine, 
first consider the case of a spin chain with periodic boundary conditions
that is at its critical point.  The system can be moved away from criticality 
by turning on a uniform magnetic field, say in the x-direction, at each site:
\begin{eqnarray}
H_B = h~ \sum_{j=1}^{L} S^x_j\ .
\end{eqnarray}
This perturbation makes the correlation length finite: 
\begin{eqnarray}
\xi_B ~\propto~ |h|^{-\nu_B}\ .
\end{eqnarray}
Explicit dimerization, breaking the symmetry of translation by 
one site, also moves the system away from criticality. For a Heisenberg 
antiferromagnet, this can be realized by the addition of a staggering term 
$R$ to the Hamiltonian:
\begin{eqnarray}
H = \sum_{j=1}^{L-1} [1 + (-1)^j R]~ \vec{S}_j \cdot \vec{S}_{j+1}\ .
\end{eqnarray}
The correlation length $\xi$ in this case scales as 
\begin{eqnarray}
\xi ~\propto~ |R|^{-\nu}\ .
\end{eqnarray}
Thus there are two independent exponents which correspond to 
these two perturbations 
of critical spin chains.  Two-parameter scaling functions can be written 
for various observables and, for a finite 
system, these involve two dimensionless variables: the ratios $L/\xi$ 
and $L/\xi_B$.  The induced dimerization, defined for now as the modulation of
the $x-x$ and $y-y$ spin-spin correlations on even versus odd links,  
\begin{equation}
\Delta = (-1)^j ~ \left[ \langle S^x_j S^x_{j+1} + S^y_j S^y_{j+1} \rangle - 
\langle S^x_{j-1} S^x_j + S^y_{j-1} S^y_j \rangle \right]\ ,
\end{equation}
is of course independent of the site index for periodic chains, and scales 
as a function of the 
chain length $L$, the field $h$, and the dimerization parameter $R$ as:
\begin{eqnarray}
\Delta(L, R, h) = {\rm sgn}(R)~ 
|R|^{\alpha_{\Delta}} ~f_{\Delta} (L |R|^{\nu}, L |h|^{\nu_B}) \ .
\end{eqnarray}
When the applied magnetic field is removed, $h=0$, and 
this expression simplifies to: 
\begin{eqnarray}
\Delta(L, R) &=& {\rm sgn}(R)~ 
|R|^{\alpha_{\Delta}} ~g_{\Delta}(L |R|^{\nu})  \nonumber \\
&\sim& L^{x_{\Delta}}~ R ~~~~ {\rm as} ~~~R \rightarrow 0,
\label{hzero}
\end{eqnarray}
where the second line follows from the fact that when the perturbation $R$ is 
very small, or equivalently when the correlation length is larger than the 
system size, the net induced dimerization must be an analytic, linear, 
function of $R$.  Therefore, for $|x| \ll 1$, the scaling function 
$g_{\Delta}(x)$ is given by:
\begin{eqnarray}
g_{\Delta}(x) = |x|^{-\alpha_{\Delta}/\nu} ~(a_1 |x|^{1/\nu} 
+ a_2 |x|^{2/\nu} + \ldots);
\end{eqnarray}
the first term yields linear dependence of 
$\Delta$ in $R$ in the $R \rightarrow 0$ limit, in agreement with 
Eq. \ref{hzero}, and the subsequent terms are higher order corrections. 
To recover the correct $L$-dependence, we must set
\begin{equation}
x_{\Delta} = {{1-\alpha_{\Delta}}\over{\nu}}\ .  
\end{equation}
The exponent $x_{\Delta}$ and the correlation length exponent $\nu$ satisfy 
the usual relation 
\begin{equation}
\nu = {{1}\over{2-x_{\Delta}}}\ .
\end{equation}

The applied magnetic field also polarizes the spins along the chain.  
The scaling form for the spin moment at each site is given by:
\begin{eqnarray}
\langle S^x \rangle = {\rm sgn}(h)~
|h|^{\alpha_B} ~f_B(L |R|^{\nu}, L |h|^{\nu_B})\ . 
\end{eqnarray}
With no applied dimerization, $R=0$, and we expect the simple power-law: 
\begin{eqnarray}
\langle S^x \rangle \sim L^{x_B} ~h ~~~~ {\rm as} ~~~h \rightarrow 0\ .
\end{eqnarray}
Therefore, $x_B = (1-\alpha_B)/\nu_B$.

\begin{figure}
\epsfxsize=4in \epsfbox{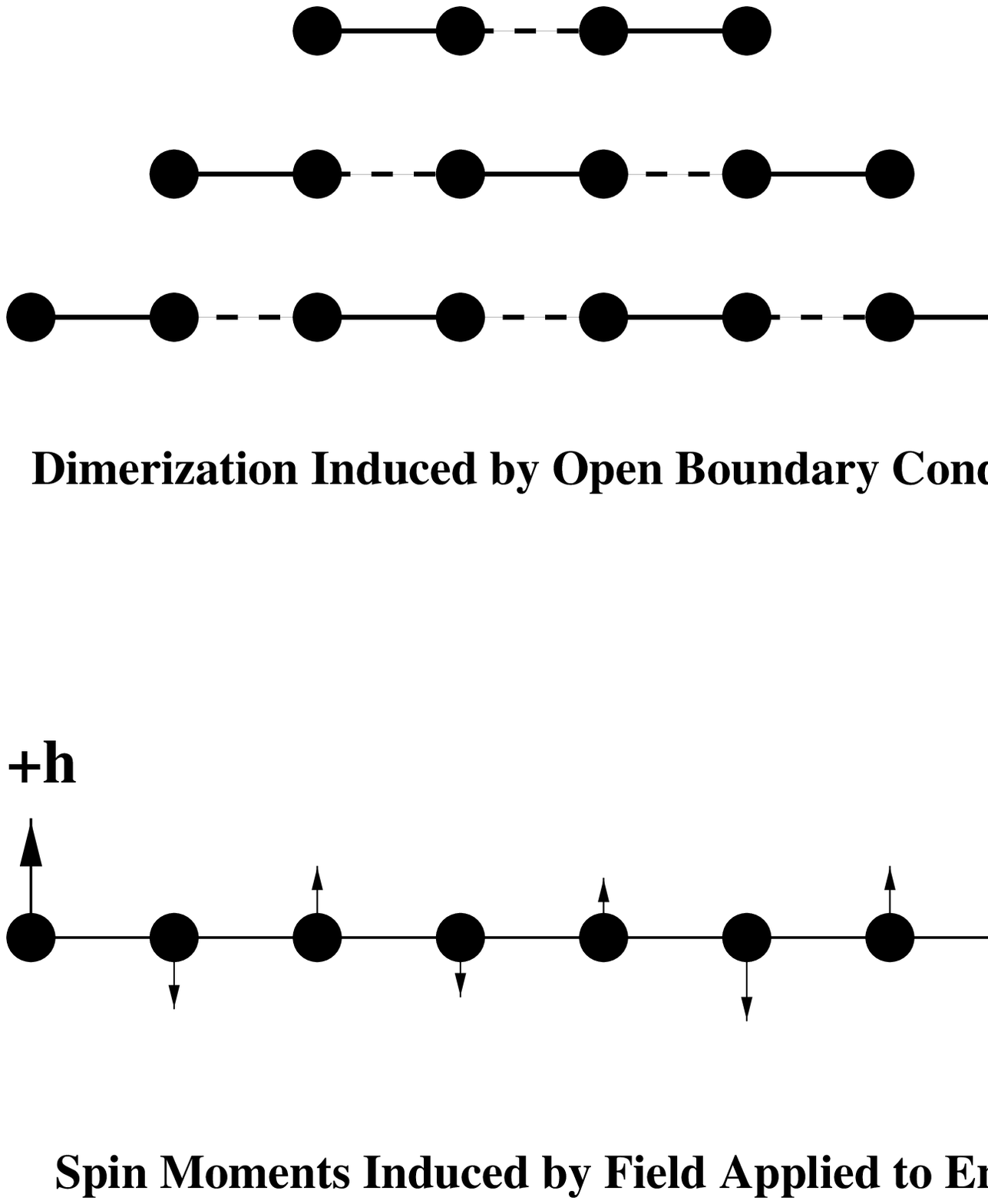}
\caption{Extraction of critical behavior from finite-size effects. 
Dimerization of the nearest-neighbor spin-spin correlation function, 
indicated here by alternating strong (solid) and weak (dashed) bonds, 
is induced by the open boundary conditions. Spin moments are induced by 
the application of a magnetic field of strength $\pm h$ to the two spins 
at the ends of the chains.}
\label{bc}
\end{figure}

Alternatively, dimerization can be induced by open boundary conditions, and we
take advantage of this fact to extract critical exponents.  
As depicted in Fig. \ref{bc}, open boundary conditions favor enhanced 
nearest-neighbor spin-spin correlations on the two outermost links.  Chains
of increasing length $L = 4, 6, 8, \ldots$ exhibit alternating patterns of
dimerization on the interior bonds.  Likewise, spin moments may be induced
in the interior of the chain by 
applying a magnetic field to the ends of the chain.  Strong applied 
edge magnetic fields completely polarize the end spins and induce non-zero
and alternating spin moments along the chain.  Alternatively, spin moments can 
be induced as before by a staggered magnetic field applied along the entire
chain.  Here however we consider only edge magnetic fields. 
We monitor the induced dimerization and spin 
moments at the center of the chain as the chain length $L$ is enlarged via 
the DMRG algorithm.  This scaling analysis is convenient because the 
relatively simple infinite-size DMRG algorithm applies to open chains 
and is most accurate at the center region of the chain where we focus our
attention.   The induced dimerization and spin moments in the 
interior of the chain show power-law scaling at the 
critical point\cite{deGennes}.  Igloi and Rieger demonstrated power-law
scaling for a variety of open boundary conditions 
(free, fixed and mixed)\cite{Igloi}.  At the critical point 
$R=0$ and $h=0$ the induced 
dimerization scales as a power-law with possible multiplicative logarithmic
corrections:
\begin{equation}
\Delta(L/2) = L^{-x_{\Delta}} ~(\ln L)^{y_{\Delta}} ~\big{(} a + \frac{b}{L} 
+ \ldots \big{)}\ .
\label{scaling}
\end{equation}
A similar expression holds for the induced spin moment at the center of the
chain,
$\langle S^x(L/2) \rangle$, with the replacement
of the exponents $x_\Delta \rightarrow x_B$ and $y_\Delta \rightarrow y_B$. 

\subsection{Infinite-Size DMRG Algorithm}
\label{subsec:dmrg}
The name ``density-matrix renormalization-group'' is something of a misnomer as
the method is most accurate away from critical points, when there is an energy 
gap for excitations.  It is helpful to think of the DMRG algorithm as 
a systematic variational approximation for the calculation of the ground
state and/or low-lying excitations, principally in one dimension.  The Hilbert
space of a quantum chain generally grows exponentially with the chain length,
and eventually must exceed available computer memory.  The DMRG algorithm
is an efficient way to truncate the Hilbert space; as the size of the space
retained can be varied (up to machine limits) it is possible to ascertain the
size of errors introduced by the truncation.  

\begin{figure}
\epsfxsize=4.5in \epsfbox{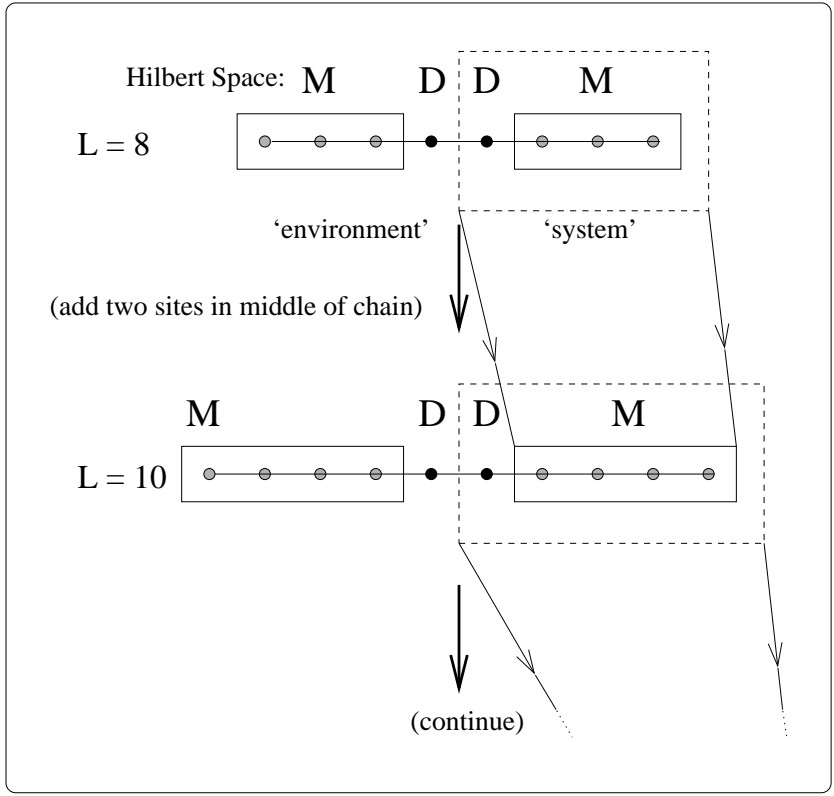}
\caption{Schematic of one iteration in the infinite-size DMRG algorithm. 
At each step, a ``new block'' is formed for each half of the chain (the 
``old block'' plus one additional site) and two more sites are added 
in the middle of the chain, increasing its length by two.}  
\label{dmrg1}
\end{figure}

For simplicity, we use the so-called ``infinite-size'' DMRG 
algorithm\cite{White}.  As the algorithm has been described in some detail
by White, we just sketch the essentials of the method.  It begins with the 
(numerically exact) diagonalization of an
open chain consisting of just four sites, each site having on-site Hilbert
space of dimension $D$.  For quantum spin chains $D= 2S + 1$, thus $D = 2$ 
for the spin-1/2 Heisenberg 
antiferromagnet.  The chain is then cut through the middle into two pieces, 
one half of which is interpreted as the ``system'' and the other half as 
the ``environment,'' the two parts combined being thought of as the entire 
``universe'' of the problem, see Fig. \ref{dmrg1}.  
At this point the reduced density matrix for the system, of size $D M \times
D M$ is constructed by performing a partial trace over the environment 
half of the chain.  It is defined by:
\begin{equation}
\rho_{ij} = \sum_{i^\prime = 1}^{DM} \Psi_{i i^\prime} \Psi_{j i^\prime},
\end{equation}
where $\Psi_{i i^\prime} = \langle i i^\prime | \Psi \rangle$ are the 
real-valued matrix elements of the eigenstate of interest 
(the ``target'' which is often the ground state) projected onto a basis of 
states labeled by unprimed Roman index $i$ which covers the system half of the 
chain and primed index $i^\prime$ which covers the environment half of the 
chain.
The eigenvalues of the reduced density matrix are real, positive, and sum up
to one; these are interpreted as probabilities.  We keep only the $M$ 
most probable
eigenstates corresponding to the largest eigenvalues, and discard the 
remaining $M (D-1)$ eigenstates.  The retained states form a new basis for
the problem.  Next, two new sites are added to the middle of the chain and the
pieces are connected, yielding a chain of size $L = 6$.  The process is then 
repeated by finding the targeted state of this chain, constructing the 
new reduced density matrix and again projecting onto the $M$ most 
probable states.  As the chain length grows in steps of two, the total
Hilbert space dimension grows by a multiplicative factor of $D^2$.  None of
the Hilbert space is thrown away until the chain grows large enough that its
Hilbert space exceeds the space that is held in reserve, 
in other words until $D^L > D^2 M^2$.
The truncation process damages the outer regions of the chain the most, and
the central region is treated most accurately.

One advantage of the method presented in this paper is that critical 
exponents are extracted from ground-state correlations only.  Excited 
states are not needed for these exponents and there 
is no need to calculate the excitation gap.  Furthermore, the finite size 
analysis described in the previous subsection takes advantage of the fact that 
the DMRG algorithm works best with open chains and treats the central 
region of the chain most accurately.  The use of the more complicated 
finite-size algorithm might yield even more accurate results.  However, we show
below that we can calculate critical exponents to an accuracy of a few percent
or better with the infinite-size algorithm.

\section{Tight Binding Model at Half-Filling}
\label{sec:tightbinding}

As a simple first illustration of our finite-size scaling method we study 
the ordinary tight binding model of spinless fermions hopping from 
site-to-site along a chain at half-filling. 
Obviously, the DMRG algorithm is not needed in this case as we can 
solve the quadratic problem exactly via a Fourier transform.  Due to 
particle-hole symmetry, at half-filling the chemical potential is zero.  The 
correlation length exponent for this system is $\nu = 1$.  A direct way to 
see this is by introducing the staggering parameter $R$ to modulate the 
amplitude of the hopping matrix elements on even versus odd links:
\begin{eqnarray}
H &=& t \sum_{j=0}^{L-1} [1 + (-1)^j R] \big{(} c^\dagger_j c_{j+1} 
+ h.c. \big{)}\ .
\label{stag-tb}
\end{eqnarray}

To diagonalize the Hamiltonian, in the case of periodic boundary conditions 
$c_0 = c_{L}$, we introduce separate fermion operators for even and odd sites 
as follows:
\begin{eqnarray}
c_{2j} = d_{2j} \nonumber \\
c_{2j-1} = e_{2j}
\label{evenodd}
\end{eqnarray}
After the Fourier transformation to momentum-space, the Hamiltonian can be 
written as:
\begin{eqnarray}
H = t \sum_k \bigg{\{} [ (1-R) + e^{2ik}(1+R) ] ~d_{k}^{\dagger} e_k + 
[ (1-R) + e^{-2ik}(1+R) ] ~e^{\dagger}_{k} d_k \bigg{\}},
\label{fourier-tb}
\end{eqnarray}
where the lattice spacing $a = 1$.  For each $k$, 
diagonalization of the $2 \times 2$ matrix yields the dispersion relation:
\begin{eqnarray}
\epsilon_k = \pm 2 t \sqrt{ 1 - (1-R^2) \sin^2(k) }.
\label{dispersion-tb}
\end{eqnarray}
At half-filling the ground state has all states with $\epsilon_k < 0$ 
occupied.  The left and right Fermi points are, respectively, 
$k_F = \pm \pi/2$.  Hence the gap $m = 2 t |R|$. 
As the correlation length $\xi \propto m^{-1} \propto |R|^{-1}$ 
we obtain $\nu = 1$.  Since $\nu^{-1} = 2 - x_{\Delta} = 1$, 
the dimerization exponent $x_\Delta = 1$.
  
\begin{figure}
\epsfxsize=5in \epsfbox{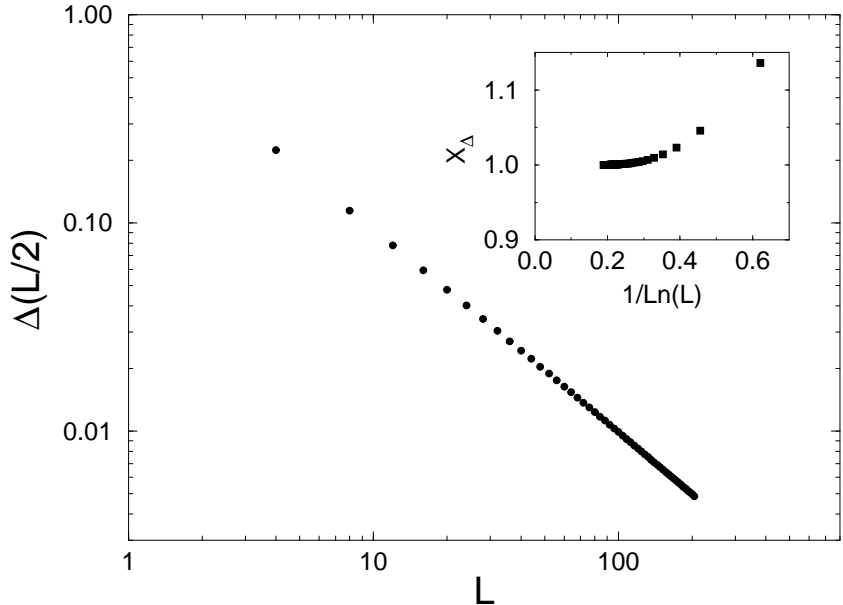}
\caption{Induced dimerization in the tight binding model.  A log-log plot 
of the dimerization at the center of the chain (Eq. \ref{dimer-tb}) is 
shown as a function of the chain length $L$.  For more complicated systems, 
the DMRG algorithm is employed to calculate $\Delta(L/2)$ numerically.
In the inset, the dimerization exponent calculated from the slope of the 
curve shown in the main graph is plotted as function of $1/\ln(L)$.  For 
small $L$ there are subleading corrections to scaling, but 
$x_{\Delta} \rightarrow 1.0$ as the chain length increases.}
\label{dim-tb}
\end{figure}

We now reproduce this result using the finite-size scaling method applied
to open chains.   We consider a finite chain of 
length $L$ with open boundary conditions and calculate the induced dimerization 
$\Delta(j) =  (-1)^j ~\langle c^\dagger_j c_{j+1} 
- c^\dagger_{j+1} c_{j+2} \rangle $ around the
chain center $j = L/2$, and extract its leading dependence on $L$. 
Open boundary conditions are imposed by using the Fourier transform
\begin{eqnarray}
c_j &=& \frac{1}{\sqrt{2 (L+1)}} \sum_{m=1}^L c_{k_m}~ 
( e^{i k_m j} - e^{-i k_m j} ), \nonumber \\
k_m &=& \frac{\pi}{L+1} m, \hskip 1cm m=1, 2, \ldots, L 
\label{ft-tb}
\end{eqnarray}
as this enforces $c_0 = c_{L+1} = 0$.  Filling all of the negative energy 
states at half-filling, the expectation value of the dimerization at $L/2$ 
can be found by straightforward calculation:  
\begin{equation}
\Delta(L/2) \propto \frac{1}{L+1} \sum_{m=\frac{L}{2}+1}^L \big{[} 
\cos[k_m (L+3)] - \cos[k_m (L+1)] \big{]}.
\label{dimer-tb}
\end{equation}
This sum can be evaluated numerically with the result that $x_{\Delta}
\rightarrow 1$ as $L \rightarrow \infty$ as shown in Fig. \ref{dim-tb}, 
in agreement with the explicit calculation for the periodic chain.  It is
also easy to show that open 
chains with an odd number of sites have vanishing induced 
dimerization at the center of the chain, 
as expected by the symmetry of reflection about the central site.

The induced density moment can likewise be obtained either directly by 
studying the effects of a staggered chemical potential $\mu_{stag}$ 
(which doubles the size of the unit cell from one to two sites and 
thus generates a gap $m = 2 |\mu_{stag}|$) or by the inclusion a local 
chemical potential $\mu$ at the two ends of the chain:
\begin{eqnarray}
H \rightarrow H  - \mu~ (c^{\dagger}_0 c_0 + c^{\dagger}_{L-1} c_{L-1})\ .
\end{eqnarray}
Again the system consists of $L$ sites, the site index running from 
$0$ to $L-1$, and there are open boundary condition at $j = 0$ and $j = L-1$. 
For large $\mu \gg 0$, the boundary condition is equivalent  
to enforcing unit occupancy at the chain ends, $n_0 = n_{L-1} = 1$. 
This boundary condition is satisfied by the Fourier transform
\begin{eqnarray}
c_j &=& \frac{1}{\sqrt{2 (L-1)}} \sum_{m=0}^{L-1} c_{k_m}~ 
( e^{i k_m j} + e^{-i k_m j} )
\end{eqnarray}
with
\begin{eqnarray}
k_m &=& \frac{\pi}{L-1} m, \hskip 1cm m=0, 1, \ldots, L-1 \ .
\end{eqnarray}
Again it is a simple exercise to calculate the occupancies.  At the chain
ends we obtain: 
$\langle c^{\dagger}_0 c_0 \rangle = \langle c^{\dagger}_{L-1} c_{L-1} \rangle 
= 1 $ in agreement with the boundary 
condition.  At the center of the chain the occupancy can be evaluated 
analytically, 
\begin{eqnarray}
\langle c^{\dagger}(L/2) c(L/2) \rangle = 
\frac{1}{L-1} \sum_{m=\frac{L}{2}}^{L-1} \left[ 1 + \cos(k_m L) \right].
\end{eqnarray}
It scales as
$\langle c^{\dagger} (L/2) c(L/2) \rangle - 1/2 \propto L^{-1}$.  Hence $\nu_B 
= x_B = 1$ in agreement with the direct calculation of these exponents.  
 
\section{Spin-$1/2$ Antiferromagnet}
\label{sec:s=half}
We next turn to the study of a richer system: spin-$1/2$ antiferromagnetic 
chains. We begin with the XY model, which can be solved exactly by a 
Jordan-Wigner mapping to the tight binding model.  We then study the 
anisotropic XXZ model.  The isotropic Heisenberg model is treated separately
as there are complicating multiplicative logarithmic corrections
to scaling at the isotropic point.

\subsection{XY model}
\label{subsec:xy}
The Hamiltonian for the spin-$1/2$ XY model, 
\begin{equation}
H = J \sum_{j=0}^{L-2} \big{[} S^x_j S^x_{j+1} + S^y_j S^y_{j+1} \big{]} \ ,
\label{h-xy}
\end{equation}  
can be written in terms of spinless fermion creation and annihilation 
operators $c_j^\dagger$ and $c_j$ via the Jordan-Wigner 
transformation\cite{Jordan}.  An up spin in the z-direction at site $i$ 
then corresponds to having the site occupied 
by a fermion, while spin down corresponds to an empty site.  The Hamiltonian 
of Eq. \ref{h-xy} is mapped to a nearest-neighbor tight binding Hamiltonian 
with $t = J$.  Based on our analysis in the previous section we can conclude
that $\nu = 1$ for the XY model. 

\begin{figure}
\epsfxsize=5in \epsfbox{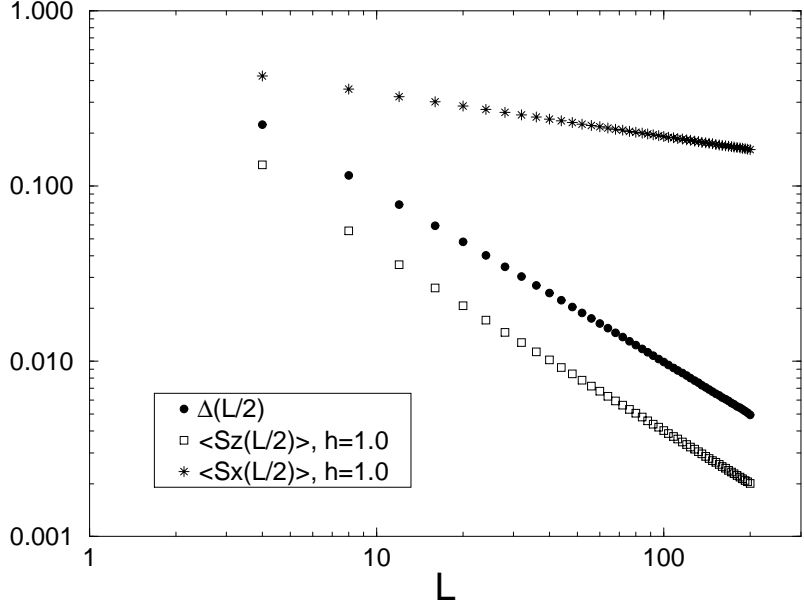}
\caption{Log-log plots of induced dimerization and induced spin moments 
in the z-direction and x-direction (with the edge magnetic field applied
in the z- and in the x-directions, respectively) 
at the center of the chain for the 
spin-$1/2$ XY model. The magnitude of the applied edge magnetic field is 
$h = 1.0$ in both cases and the number of block states kept is $M = 128$.}
\label{xy}
\end{figure}

Fig. \ref{xy} presents our DMRG results for the induced dimerization and 
induced spin moments, in the x- and in the z-directions, at the center of the 
chain as a function of the chain length, $L$.
The exponents are obtained from the slopes of the curves shown in Fig. 
\ref{xy}.  The induced dimerization exponent for $\Delta(L/2)$ is 
close to $1$ ($x_{\Delta} = 0.99 \pm 0.01$) as expected from the relation 
$\nu = 1/(2-x_{\Delta})$.  The slope of the log-log plot of the induced 
spin moment in the z-direction is also close to $1$ 
($x_B = 1.01 \pm 0.02$). This result is also expected since it is equivalent 
to the exponent for the induced density moment in the tight binding model as
discussed in the previous section. 
In the case of the induced spin moment in the x-direction, the exponent is 
$0.248 \pm 0.003$.  This value compares well with the exact number of $1/4$ 
as derived in the next section.

\subsection{XXZ model}
\label{subsec:xxz}
Next consider the nearest-neighbor, spin-1/2 XXZ Heisenberg antiferromagnet: 
\begin{equation}
H = J \sum_{j=0}^{L-2} \big{[} S_j^x S_{j+1}^x + S_j^y S_{j+1}^y + 
\gamma S_j^z S_{j+1}^z \big{]}\ .
\label{Heisenberg}
\end{equation}
Anisotropy in the coupling between the z-components of the spins may be 
varied by changing $\gamma$.
Performing the Jordan-Wigner transformation, the XY terms again
yield the tight binding Hamiltonian.  Low-energy excitations therefore occur
near the two Fermi points at $k = \pm \pi / 2a$. 
We may treat the non-Gaussian $\gamma$ term as a perturbation and focus on 
excitations around these Fermi points by defining left and right moving 
low-energy quasiparticles.  Taking the continuum limit and keeping only 
the low-energy modes, the 
tight binding term is then effectively described by the massless 
fermions.  The $S_j^z S_{j+1}^z$ term is quartic in the fermion 
operators.  Integrating out the high-energy modes, it will renormalize the 
fermion velocity and also contain interaction terms. 

We then implement Abelian bosonization, with UV cutoff $\alpha$.  The 
effective Hamiltonian is a sine-Gordon model (a derivation can be found in 
Ref. 49):
\begin{equation}
H = H_0 - \frac{y_{\phi}}{2 \pi \alpha^2} \int dx 
\cos[\sqrt{8 \pi} \phi(x)]
\label{sine-Gordon}
\end{equation}
where
\begin{equation}
H_0 = u \int dx \bigg{[} K \Pi^2 + \frac{(\partial_x \phi)^2}{K} \bigg{]}.
\label{freeboson1}
\end{equation}
Here $u = 2 J a = 2 a$ is the bare Fermi velocity 
and the constant $K \equiv 1 + y_0/2$ depends on the anisotropy $\gamma$. The 
XY limit correspond to $y_{\phi} = 0$.

\begin{figure}
\epsfxsize=5in \epsfbox{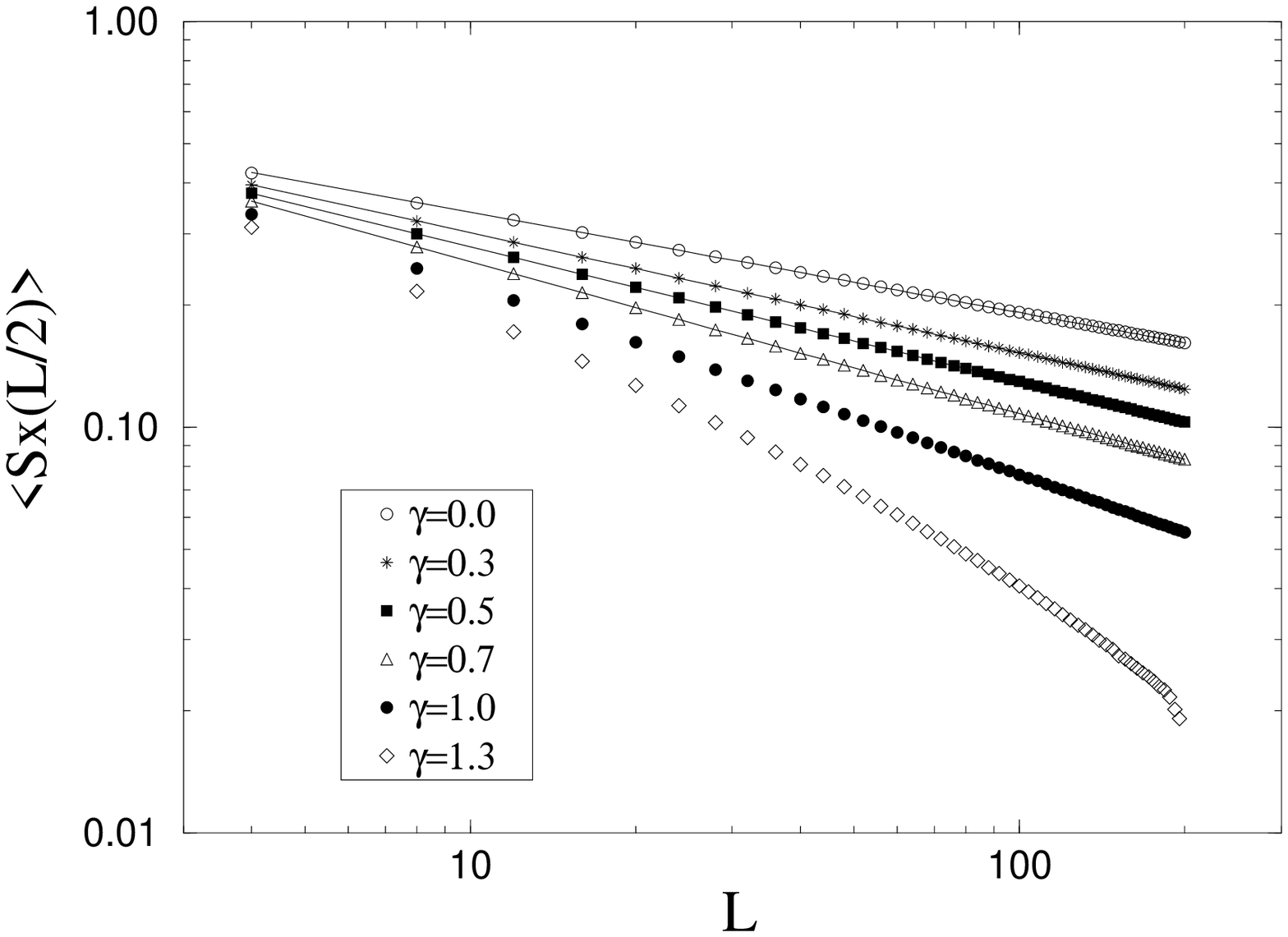}
\epsfxsize=5in \epsfbox{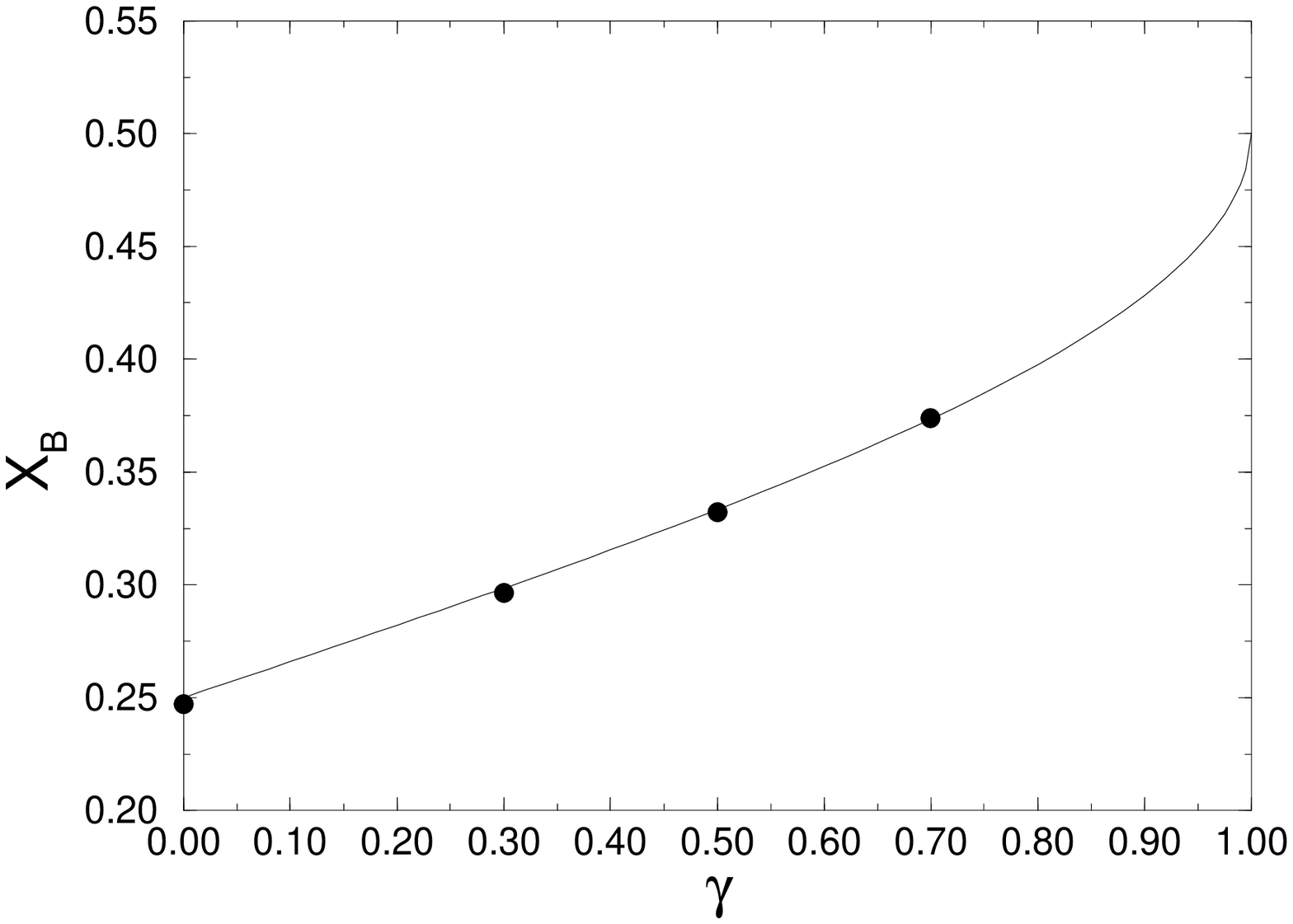}
\caption{Spin-$1/2$ XXZ Heisenberg antiferromagnetic chain. (a) DMRG 
results ($M=128$ and $h=1.0$) for the induced spin 
moment in the x-direction at center of chain for different values of 
anisotropy $\gamma$, showing power law decay for $\gamma \le 1$ and 
exponential decay for $\gamma > 1$ and (b) corresponding exponent 
$x_B(\gamma)$ for $\gamma < 1$.  The solid line is the exact value 
$\pi R^2(\gamma)$. The exponents are obtained by fitting the curves in (a) 
to the form of Eq. \ref{scaling} with $y_B = 0$ since there are no 
logarithmic corrections away from the isotropic point.}
\label{s12_sxgamma}
\end{figure}

The long distance behavior of the staggered part of $S^z$ and $S^-$ are 
given in terms of the boson fields as:
\begin{eqnarray}
S^z(x) &\approx& (-1)^{x/\alpha} \cos [ \phi(x)/R ] \nonumber \\
S^-(x) &\approx& (-1)^{x/\alpha} e^{i 2 \pi R \tilde{\phi}(x)}
\label{spinops}
\end{eqnarray}
where the radius $R$ is given by\cite{Affleck}
\begin{eqnarray}
R=\sqrt{ \frac{1}{2\pi} - \frac{\cos^{-1} \gamma}{2 \pi^2} } .
\label{R}
\end{eqnarray}
First consider the anisotropic case $\gamma \neq 1$.  The isotropic 
case has logarithmic corrections to scaling that are dealt with in the next 
section.  For $\gamma > 1$ the interaction term is relevant and the system is 
gapped, and in the Ising universality class.  Indeed, in the limit
$\gamma \rightarrow \infty$ it is the Ising model. 
For $\gamma < 1$ the interaction term is irrelevant, the system is gapless 
and $\Delta(L/2)$ and 
$\langle S^x(L/2) \rangle$ should exhibit power law decay, 
with no log corrections as there are no marginal operators. 
The log-log plots of Fig. \ref{s12_sxgamma} (a) show the induced spin moment 
in the x-direction at the chain center for different values of the 
anisotropy $\gamma$. The edge magnetic field in the x-direction is 
fixed, $h = 1.0$. As expected, for 
$\gamma > 1$ there is exponential decay and in the cases $\gamma < 1$ 
the exponents $x_B(\gamma)$ are found by fitting the curves in 
Fig. \ref{s12_sxgamma} (a) to the form of Eq. \ref{scaling}.  The exponents 
$y_B$ are set equal to zero, the higher order corrections are included 
and give very small deviations from a simple linear fit.  In 
Fig. \ref{s12_sxgamma} (b) the exponents $x_B(\gamma)$ are compared to 
the exact value $x_B(\gamma) = \pi R^2(\gamma)$ obtained by 
Affleck\cite{edgefield}.  Agreement
is found at the percent level.  Affleck derived the exponent as follows.  
The edge magnetic field in the x-direction applied at $j=0$ corresponds to 
a term
\begin{equation}
H_B = - h S^x(0) = -{\rm constant} \times h \cos[\sqrt{2 \pi} \tilde{\phi}(0)]
\label{edgefield}
\end{equation}
in the Hamiltonian.  For sufficiently large $h$ the energy is minimized by
setting
\begin{equation}
\tilde{\phi}(0) = 0 \Longrightarrow \phi_R(0) = \phi_L(0) \ .
\label{BC}
\end{equation}
Regarding $\phi_R$ as an analytic continuation of $\phi_L$, we may identify
\begin{equation}
\phi_R(x) = \phi_L(-x) \ .
\label{BCgen}
\end{equation}
Using this boundary condition, the induced spin moment is given by
\begin{eqnarray}
\langle S^x(j) \rangle \approx (-1)^{j/\alpha} 
\langle e^{i 2 \pi R \phi_L(j)} e^{-i 2 \pi R \phi_L(-j)} \rangle 
\approx \frac{(-1)^{j/\alpha}}{(2j)^{\pi R^2(\gamma)}} \ .
\end{eqnarray}
For the XY model ($\gamma = 0$), the induced spin moment in the x-direction 
therefore decays with exponent $\pi R^2(0) = 1/4$.

\begin{figure}
\epsfxsize=5in \epsfbox{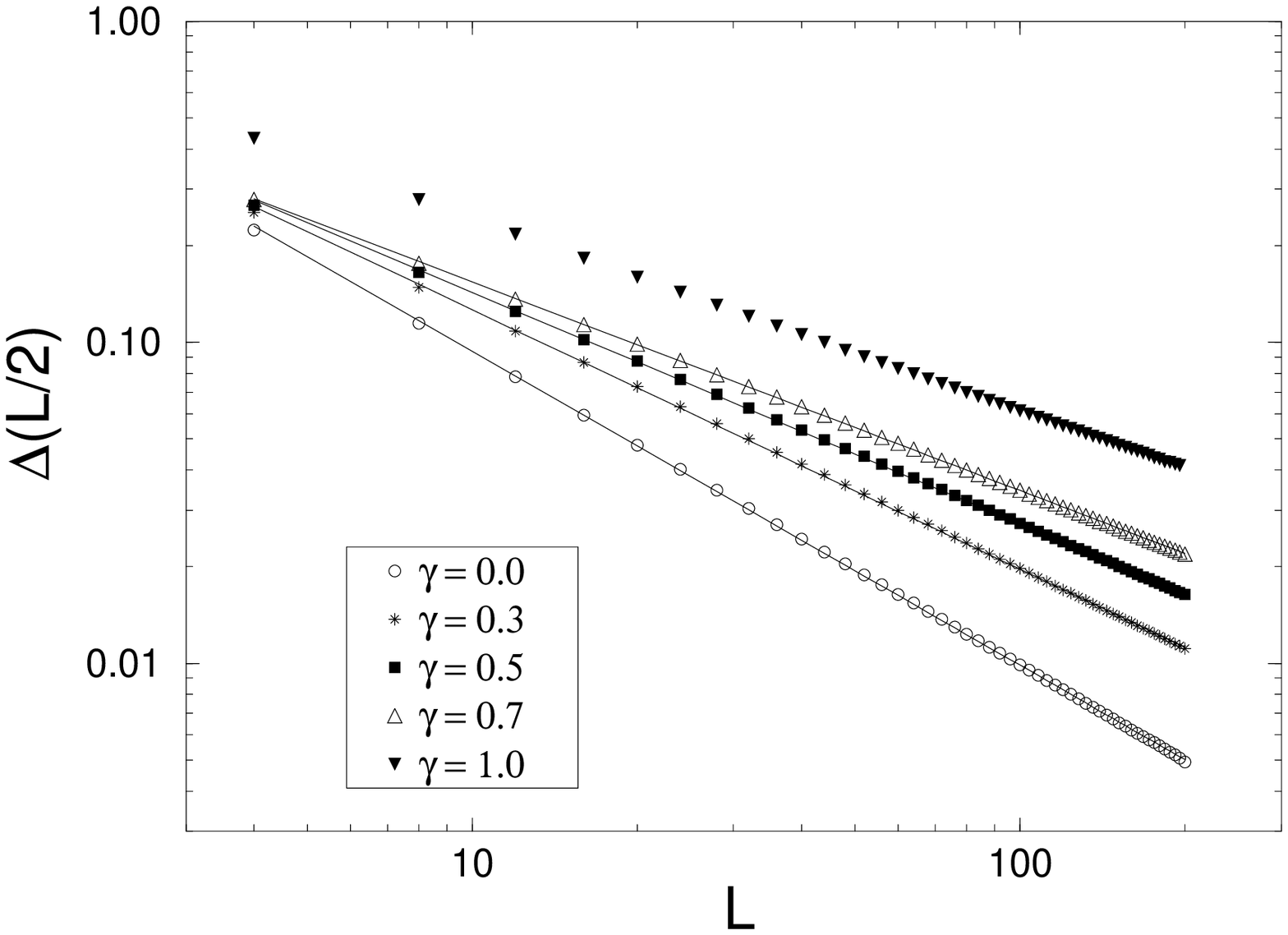}
\epsfxsize=5in \epsfbox{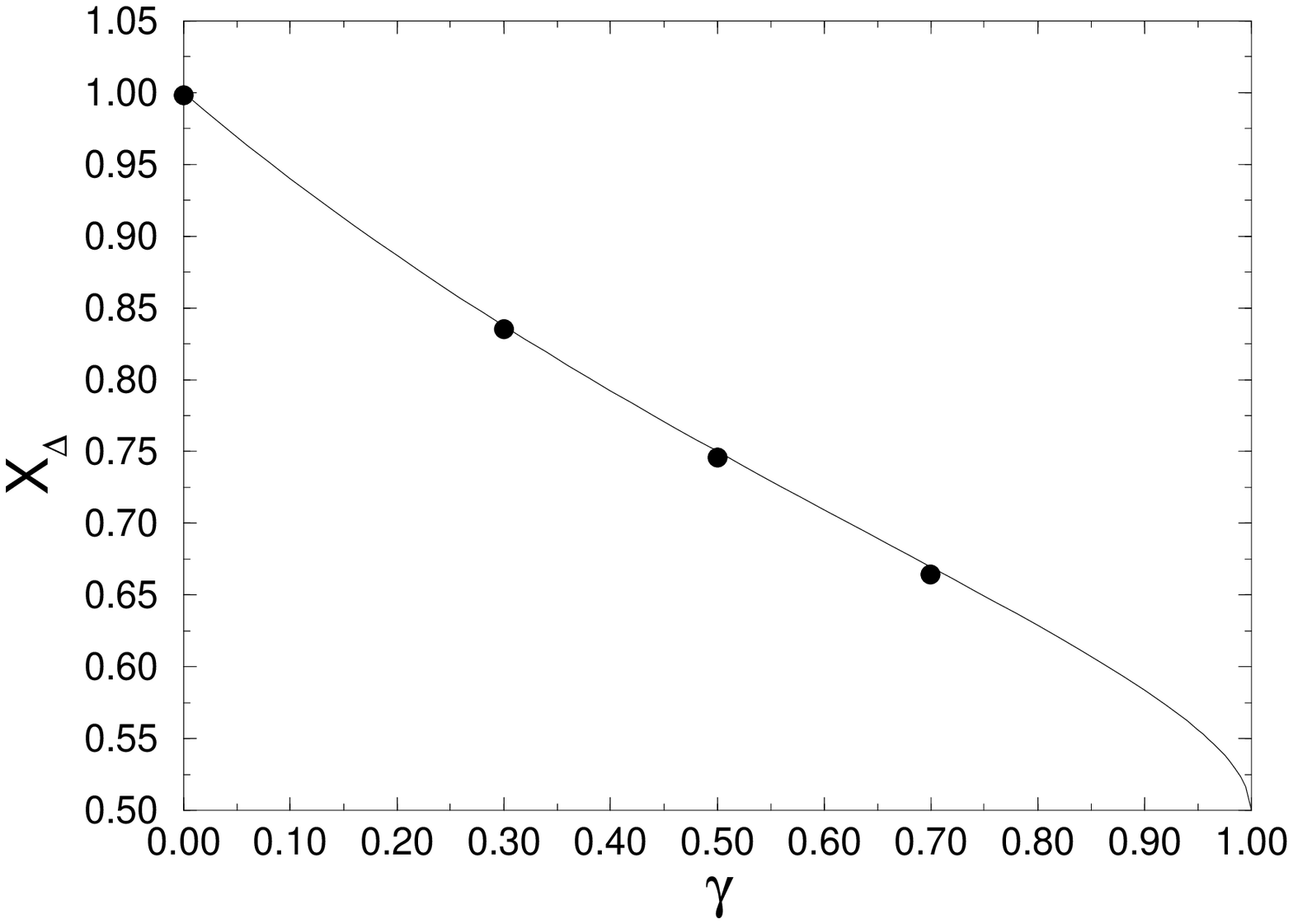}
\caption{Spin-$1/2$ XXZ Heisenberg antiferromagnetic chain. (a) DMRG 
results ($M=128$) for the induced dimerization at center of chain 
for different values of anisotropy $\gamma$ and (b) corresponding exponent 
$x_{\Delta}(\gamma)$, obtained by fitting to Eq. \ref{scaling} with 
$y_{\Delta} = 0$. The solid line is the exact value $1/4 \pi R^2(\gamma)$.}
\label{s12_dimgamma}
\end{figure}

A log-log plot of the induced dimerization at the center of 
the chain for various values of the anisotropy $\gamma$ is shown in 
Fig. \ref{s12_dimgamma} (a).  The free boundary condition at the chain
ends corresponds to setting:
\begin{eqnarray}
\vec{S}_0 = \vec{S}_{L+1} = 0\ .
\end{eqnarray}
This condition translates to $\phi_R(x) = - \phi_L(-x) + \pi R$ in terms of 
the boson fields\cite{Eggert} which yields:
\begin{eqnarray}
\Delta(j) ~ \approx ~ (-1)^{j/\alpha} \langle 
\cos[\phi(j)/R] \rangle 
~ \approx ~ \frac{(-1)^{j/\alpha}}{(2j)^{1/4 \pi R^2(\gamma)}}.
\end{eqnarray} 
In Fig. \ref{s12_dimgamma} (b) the exponents obtained from the slopes of 
the curves in Fig. \ref{s12_dimgamma} (a) are plotted against the exact 
values $x_{\Delta} = 1/4 \pi R^2(\gamma)$.  Again agreement is found 
at the percent level.

\begin{figure}
\epsfxsize=5in \epsfbox{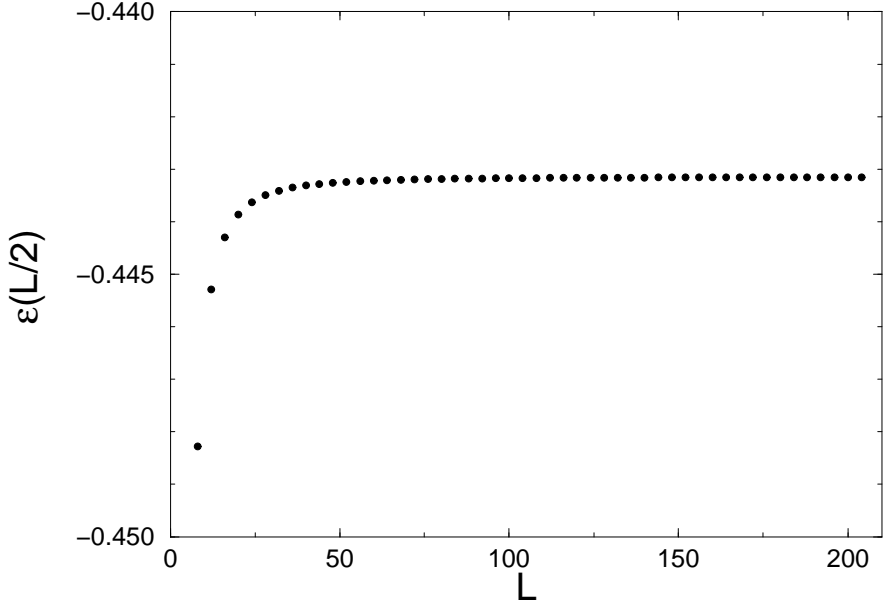} 
\caption{Sum of the two central bonds for the spin-$1/2$ XXZ Heisenberg 
antiferromagnetic chain at the isotropic point $\gamma=1$.  The block
size is $M=256$.}
\label{s12sum}
\end{figure}

\begin{figure}
\epsfxsize=5in \epsfbox{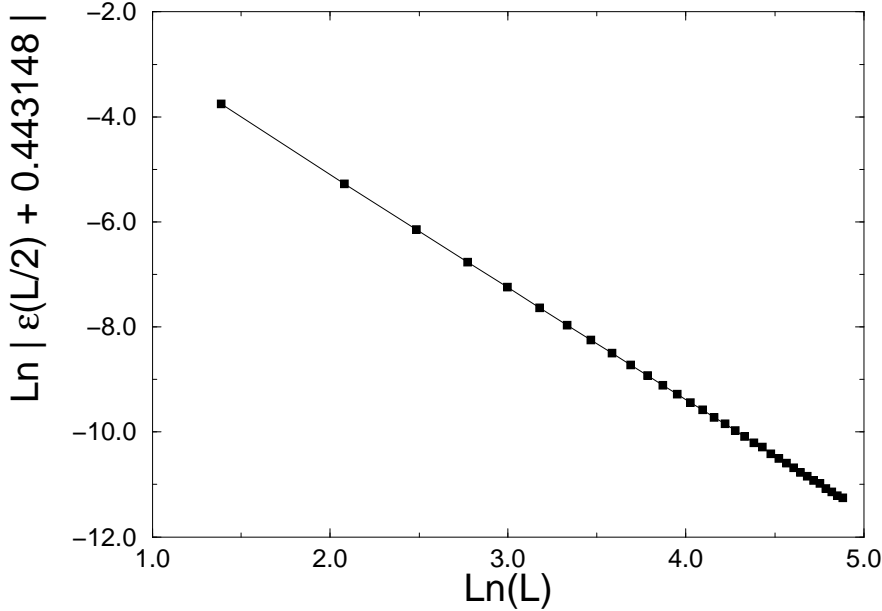}
\caption{Log-log plot of $\epsilon(L/2)$ after subtracting the 
extrapolated value $\epsilon(\infty) = -0.443148$.  The fit is to a straight
line of slope $2.14$.}
\label{s12sumexp}
\end{figure}

Another quantity of interest is the sum, instead of the difference, of 
the spin-spin correlation function on adjacent bonds near the center of the 
chain:
\begin{eqnarray}
\epsilon(L/2) ~ \equiv ~ \frac{1}{2} ~ \big{(} ~ \langle \vec{S}_{L/2} \cdot 
\vec{S}_{L/2+1} \rangle + \langle \vec{S}_{L/2-1} \vec{S}_{L/2} \rangle ~ 
\big{)} ~ ,
\end{eqnarray}
which at $\gamma = 1$ equals the energy density per bond and therefore 
does not vanish in the thermodynamic limit. Fig. \ref{s12sum} is a plot of 
$\epsilon(L/2)$ as a function of the system size $L$ at the isotropic 
point $\gamma=1$. 
As expected, this quantity approaches a constant value $\epsilon(\infty)$ 
in the thermodynamic limit.  After subtracting the extrapolated value at 
$L \rightarrow \infty$, $\epsilon(L/2)$ too exhibits power law decay 
of the form of Eq. \ref{scaling}.  The constant $\epsilon(\infty)$ can be 
found by an iteration process.  Starting with an initial value for 
$\epsilon(\infty)$ obtained from a rough extrapolation of the curve in 
Fig. \ref{s12sum}, we fit the subtracted value 
$\epsilon(L/2) - \epsilon(\infty)$ to a power-law form.  The extrapolated 
value $\epsilon(\infty)$ is then adjusted slightly until an optimal fit to 
a pure power law is attained.  The extrapolated value found this way is 
$\epsilon(\infty) = -0.443148$ and Fig. \ref{s12sumexp} shows the power law 
behavior of the subtracted quantity.
We obtain an exponent of $2.1 \pm 0.1$ in the scaling of 
$\epsilon(L/2)-\epsilon(\infty)$.  This is as expected from the 
linear dispersion relation of Heisenberg antiferromagnets: in a 
Lorentz-invariant theory the energy density operator has dimension 2.
 
The DMRG result for the energy per bond is extremely accurate and can be 
compared with the exact value obtained from the Bethe ansatz 
solution\cite{Baxter} of $~\epsilon = 1/4 - \ln 2 = -0.44314718 $.  It 
is crucial to note that the open boundary 
conditions induce staggering in the strength of the bonds along the chain. To 
eliminate this effect, the energy per bond must be calculated as the average 
of the bond energy from two consecutive bonds at the center of the chain.  
Suggestions that infinite-size DMRG results for the center region of the 
chain are not very accurate\cite{Tasaki} appear to have failed to take this 
effect into account.  We have also checked our results at different 
anisotropies. For the XY case ($\gamma=0.0$), we obtain $\epsilon(\infty) = 
-0.318310$ extrapolating from chains up to $L = 200$ and 
$M = 128$ and the exact result\cite{Baxter} is $-1/\pi = -0.3183099$.

\subsection{Logarithmic Corrections to Scaling}
\label{subsec:log}
In the isotropic XXX limit, the interaction $\cos[\sqrt{8 \pi} \phi(x)]$ 
in the low-energy effective Hamiltonian Eq. \ref{sine-Gordon} becomes 
marginal and can generate multiplicative logarithmic corrections to 
scaling.  In this section we calculate its effect on the 
scaling of the induced spin moment $\langle S^x(L/2) \rangle$ when an edge 
magnetic field $H_B$ in the x-direction is applied.  Cancellations occur and 
in this case there are no multiplicative $\ln(L)$ corrections.  As a 
practical matter, the cancellation of the logarithmic corrections means 
that numerical calculations of the exponent $x_B$ are particularly precise.  
We note that finite-size scaling of the spin-spin correlation function 
has been previously calculated for a spin-$1/2$ chain with periodic boundary 
conditions\cite{Barzykin,exact}.  
 
The coupling constants in the sine-Gordon Hamiltonian (Eq. \ref{sine-Gordon}) 
renormalize under a change of the ultraviolet cutoff $\alpha \rightarrow 
\alpha e^l$ according to the renormalization group equations\cite{Giamarchi}:
\begin{eqnarray}
\frac{dy_0}{dl} &=& -y_{\phi}^2(l), \nonumber \\
\frac{dy_{\phi}}{dl} &=& -y_{\phi}(l) y_0(l). 
\label{RG}
\end{eqnarray}
As noted in the previous section, a large edge magnetic field applied at 
$x=0$ in the x-direction enforces the boundary condition 
$\phi_R(x) = \phi_L(-x)$ (Eq. \ref{BCgen}). Thus
\begin{equation}
\langle S^x(x) \rangle \sim (-1)^{x/a} \langle 
\cos[\sqrt{2 \pi} \tilde{\phi}(x)] \rangle \sim \langle 
e^{i \sqrt{2\pi} \phi_L (x)} e^{-i \sqrt{2\pi} \phi_L (-x)} \rangle. 
\label{sx}
\end{equation}
For the free theory, which corresponds to the XY model 
$y_{\phi} = 0$, the induced spin moment is simply
\begin{equation}
\langle S^x(x) \rangle_0 \sim \exp \big{[} -K U_L (2x) \big{]}
\label{sx0}
\end{equation}
where
\begin{equation}
U_L(x) = \frac{1}{2} \ln ( \frac{\alpha + i x}{\alpha} ).
\label{U_L}
\end{equation}
But in the general XXZ case we ascertain the effect of the marginal operator
by following a procedure similar to one developed by Giamarchi and 
Schulz\cite{Giamarchi} who calculated correlation functions for finite
{\it periodic} chains.  We first define the function:
\begin{eqnarray}
F(x) \equiv e^{K U_L(2x)} \langle S^x(x) \rangle\ .
\label{fx}
\end{eqnarray}
At the XY point $y_{\phi} = 0$ clearly $F(x) = 1$.  For small $x$, 
an expansion of $F$ in powers of $y_{\phi}$ converges, and 
for sufficiently small coupling $y_{\phi}$, $F(x) \sim 1$. Upon rescaling, 
the function $F(x)$ also depends on the new length scale and on the 
rescaled coupling constants $y_0(l)$ and $y_{\phi}(l)$. By an 
argument similar to the one employed by Kosterlitz\cite{Kosterlitz}, the 
effect of rescaling $\alpha \rightarrow e^l ~\alpha$ is:
\begin{eqnarray}
F(x, \alpha e^l, y(l)) = I(dl, y(l)) F(x, \alpha e^{l+dl}, y(l+dl))\ ,
\label{kost}
\end{eqnarray}
where $y(l)$ denotes all the couplings as function of the scaling parameter 
$l$. The rescaled short distance cutoff is then $\alpha(l) = e^l ~\alpha$, 
where $\alpha$ is the initial cutoff.  Rescaling can be repeated 
until $\alpha(l) \sim x$, at which point we have:
\begin{eqnarray}
F(x, x, y( \ln (x/\alpha))) = O(1)\ .
\end{eqnarray}
The contributions to the function $F$ from repeated rescalings,  
until $\alpha(l)$ reaches $x$, can be written explicitly as:
\begin{eqnarray}
F(x, \alpha, y(\alpha)) = \prod_{l=0}^{l=\ln (x/\alpha)} I(dl, y(l)) = 
\exp \bigg{\{} 
\int_0^{\ln(x/\alpha} \ln \big{[} I(dl, y(l)) \big{]} dl \bigg{\}} \ .
\label{rescaling-f}
\end{eqnarray}

\begin{figure}
\epsfxsize=5in \epsfbox{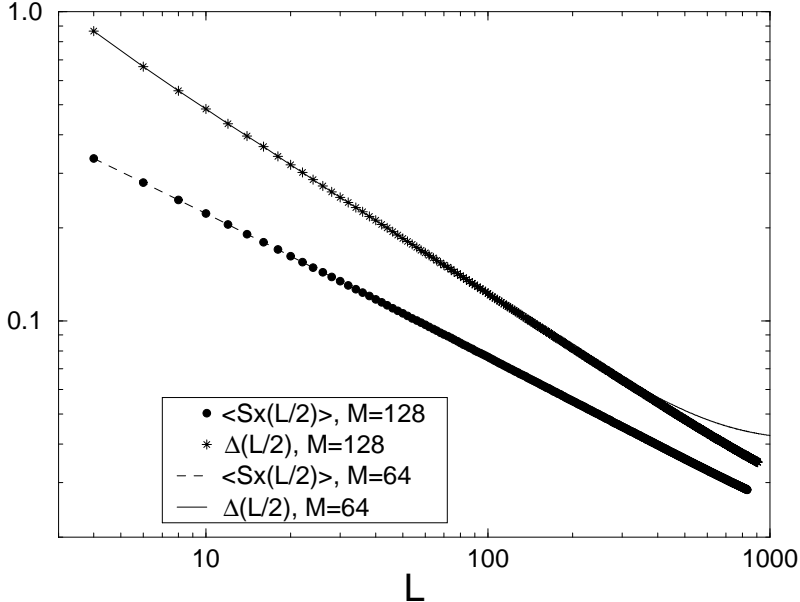}
\caption{Log-log plot of the induced spin moment ($h=1.0$) and 
induced dimerization at center of chain for 
the isotropic $S = 1/2$ Heisenberg antiferromagnet.  Data obtained using 
block Hilbert space size of $M = 128$ and $M = 64$ are plotted for 
comparison.  Note the slight curvature in $\Delta(L/2)$ which signals
the existence of multiplicative logarithmic corrections to scaling.  The
log is absent in the plot of $\langle S^x(L/2) \rangle$.  The discrepancy 
in the induced dimerization at large chain length is due to 
the truncation of the block Hilbert space.}
\label{fig-xxx}
\end{figure}

We proceed to calculate the function $I$. 
First we expand $\langle S^x(x) \rangle$ in powers of $y_{\phi}$, writing it 
in terms of averages with respect to the free Hamiltonian, 
\begin{eqnarray}
\langle S^x(x) \rangle &\sim& e^{-K U_L(2x)} + 
\frac{y_{\phi}}{2 \pi \alpha^2} \int d^2 x' 
\langle S^x(x) \cos[\sqrt{8 \pi} \phi(x')] \rangle_0 + \nonumber \\ & & 
\frac{1}{2} \big{(} \frac{y_{\phi}}{2 \pi \alpha^2} \big{)}^2 
\int d^2 x_1 \int d^2 x_2
\langle S^x(x) 
\cos[\sqrt{8 \pi} \phi(x_1)] \cos[\sqrt{8 \pi} \phi(x_2)] \rangle_0 
+ \ldots 
\label{expansion}
\end{eqnarray}
The averages are given by
\begin{eqnarray}
\langle S^x(x) \cos[\sqrt{8 \pi} \phi(x')] \rangle_0 = 0 \nonumber
\end{eqnarray}
and
\begin{eqnarray}
\langle S^x(x) 
\cos[\sqrt{8 \pi} \phi(x_1)] && \cos[\sqrt{8 \pi} \phi(x_2)] \rangle_0 
\sim \nonumber \\
&& \frac{1}{2} \exp \big{[} -K U_L(2x) + 4 K U_L(2x_1) + 4 K U_L(2x_2) - 
4 K U(x_1+x_2) - 4 K U(x_1-x_2) \big{]} 
\end{eqnarray}
The $O(y_{\phi}^2)$ term can be simplified by assuming that the main 
contribution comes from configurations where $x_1$ and $x_2$ are very 
close to each other\cite{Nelson,Giamarchi}.  Introducing new integration 
variables
\begin{eqnarray}
r &\equiv& x_1 - x_2 \nonumber \\
R &\equiv& \frac{x_1+x_2}{2} 
\end{eqnarray}
and expanding $U_L$ in powers of $r$, which is assumed to be small,
\begin{eqnarray}
U_L(2 x_1) = U_L(2R + r) = U_L(2R) + r ~\partial_R ~U_L(2R) + \ldots \ , 
\nonumber \\
U_L(2 x_2) = U_L(2R - r) = U_L(2R) - r ~\partial_R ~U_L(2R) + \ldots \ ,  
\end{eqnarray}
we obtain the average
\begin{eqnarray}
\langle S^x(x) \cos[\sqrt{8 \pi} \phi(x_1)] \cos[\sqrt{8 \pi} 
\phi(x_2)] \rangle_0 \sim \frac{1}{2} \exp \left[ 
- K U_L(2x) - 4 K U(r) \right]\ .
\end{eqnarray}
The dependence on $R$ cancels out.  The expansion Eq. \ref{expansion} becomes
\begin{eqnarray}
\langle S^x(x) \rangle &\sim& e^{-K U_L(2x)} \bigg{[} 1 + 
\frac{y_{\phi}^2 \Omega}{4 \alpha^2} 
\int_{\alpha} dr e^{-4K U(r)} \bigg{]} \ ,
\end{eqnarray}
where $\Omega \equiv \int dR$ is a measure of the linear size of the system.
Next consider the effect of rescaling $\alpha^{\prime} = \alpha e^{dl}$, 
where $dl$ is infinitesimal.  Using
\begin{eqnarray}
\int_{\alpha}^{\infty} dx = \int_{\alpha}^{\alpha^{\prime}} dx + 
\int_{\alpha^{\prime}}^{\infty} dx \ \ ,
\end{eqnarray}
we obtain: 
\begin{equation}
\langle S^x(x) \rangle \sim e^{-K U_L(2x)} \bigg{[} 1 + 
\frac{y_{\phi}^2}{4 \alpha^2} dl + \frac{y_{\phi}^{\prime 2}}
{4 \alpha^{\prime 2}} \int_{\alpha^\prime} dr e^{-4K U(r)} \bigg{]}
\label{rescaled}
\end{equation}
Matching this result with Eq. \ref{kost}, we find:
\begin{equation}
I(dl, y_0(l), y_{\phi}(l)) \sim \exp \bigg{[} \frac{y_{\phi}^2 (l)}
{4 \alpha^2} dl \bigg{]}\ ,
\end{equation}
hence from Eq. \ref{rescaling-f} and Eq. \ref{fx}, we have
\begin{eqnarray}
\langle S^x(x) \rangle  \sim  \exp \bigg{\{} - K U_L(2x) + 
\int_0^{\ln(x/\alpha)} \frac{y_{\phi}^2 (l)}{4 \alpha^2} dl \bigg{\}}\ .
\end{eqnarray}
Using the RG equations (Eq. \ref{RG}), the solution at large $l$ is 
$y_{\phi}(l) \sim 1/l$ and
\begin{eqnarray}
\langle S^x(x) \rangle \sim \bigg{(} \frac{x}{\alpha} \bigg{)}^{-1/2} \exp 
\bigg{\{} \int_0^{\ln(x/\alpha)} dl \big{[} ~O(\frac{1}{l^2})~ 
\big{]} \bigg{\}}\ .
\label{nolog}
\end{eqnarray}
There are no multiplicative $\ln(x)$ corrections, as these would require 
terms of order $O(1/l)$ in the integrand inside the exponential in Eq. 
\ref{nolog}.  In our calculation, $O(1/l)$ 
terms do not appear, only $O(1/l^2)$ and higher-order terms.  As a check, we 
can repeat the same procedure for $\langle S^z(x) \rangle$, with the edge 
field now oriented in the z-direction.  Of course this should give the same 
result since the system is isotropic, but as the Jordan-Wigner transformation
picks the z-direction as the spin quantization axis, 
the equivalence is not obvious, and the check is non-trivial.  
Again, explicit calculation shows that $O(1/l)$ terms do not arise. 
This result is in reasonable agreement with our numerical results.  Fitting 
the DMRG data (see Fig. \ref{fig-xxx}) to 
the form Eq. \ref{scaling}, we obtain $x_B = 0.485 \pm 0.01$ with a small 
non-zero value for the log exponent $y_B = 0.06 \pm 0.01$.  By contrast, in
the case of the induced dimerization we obtain $x_{\Delta} = 0.57 \pm 0.01$ 
and $y_{\Delta} = 0.10 \pm 0.05$. The error was estimated from deviations 
obtained by fitting the $M = 128$ data over different ranges of L 
($4 \le L \le 600$) and by comparison with $M = 64$ data ($4 \le L \le 300$). 
Results are systematically improved by increasing the value of $M$.  

Finite-size scaling behavior for the XXX model with open boundary 
conditions\cite{Hikihara} and periodic boundary conditions\cite{Hallberg} 
were obtained from DMRG calculations of ground state energies and 
correlation functions $\langle S^z(x) S^z(x+r) \rangle$ for different 
system sizes and separations $r$. In our approach, critical exponents 
are extracted from expectation values at the center of the chain only. 
The chain size is increased via the infinite-size DMRG method. It is also 
advantageous to extract power law exponents when there are no logarithmic 
corrections.

\subsection{Conformal Anomaly}
\label{subsec:c-half}
Finally, we may calculate the value of the conformal anomaly, $c$.  We note 
that the central charge of the RSOS model\cite{Sierra} and of the spin-$3/2$ 
Heisenberg chain\cite{Hallberg}, which is in the same universality class 
as the spin-$1/2$ chain, have previously been obtained using the DMRG.  The
conformal anomaly can be extracted by finding the coefficient of the $1/L$ 
finite-size correction to the free
energy, equivalent at zero temperature to the ground state energy.  We fit 
the ground state energy $E_0(L)$ to the following form:
\begin{equation}
E_0(L) = A L + B + C / L + \ldots
\end{equation} 
The extensive contribution, proportional to $A$, and the constant term $B$
are non-universal.  At the isotropic point $\gamma = 1$ our results for 
the case of blocksize $M = 128$ and for chain lengths in the 
range $30 \leq L \leq 100$ 
yield $C \approx -0.323$.  To relate this coefficient to the conformal 
anomaly we must normalize it by dividing by the speed of low-lying 
excitations, $v$.  The speed can be obtained by extrapolation to the 
thermodynamic limit of the gap to the lowest-lying excitation multiplied 
by the chain length: 
\begin{equation}
v = \lim_{L \rightarrow \infty}~ {{{\rm Gap}(L) \times L}\over{\pi}}\ . 
\end{equation}
We find $v = 2.44$.  Now for open boundary conditions\cite{Cardy}, 
\begin{equation}
c = {{-24 C}\over{\pi v}} \approx 1.01\ .
\end{equation}
This compares well with the value of $c = 1$ appropriate for a single 
boson or the pair of left and right moving fermions. 

\section{Spin-$1$ chain}
\label{sec:s=1}
\begin{figure}
\epsfxsize=4in \epsfbox{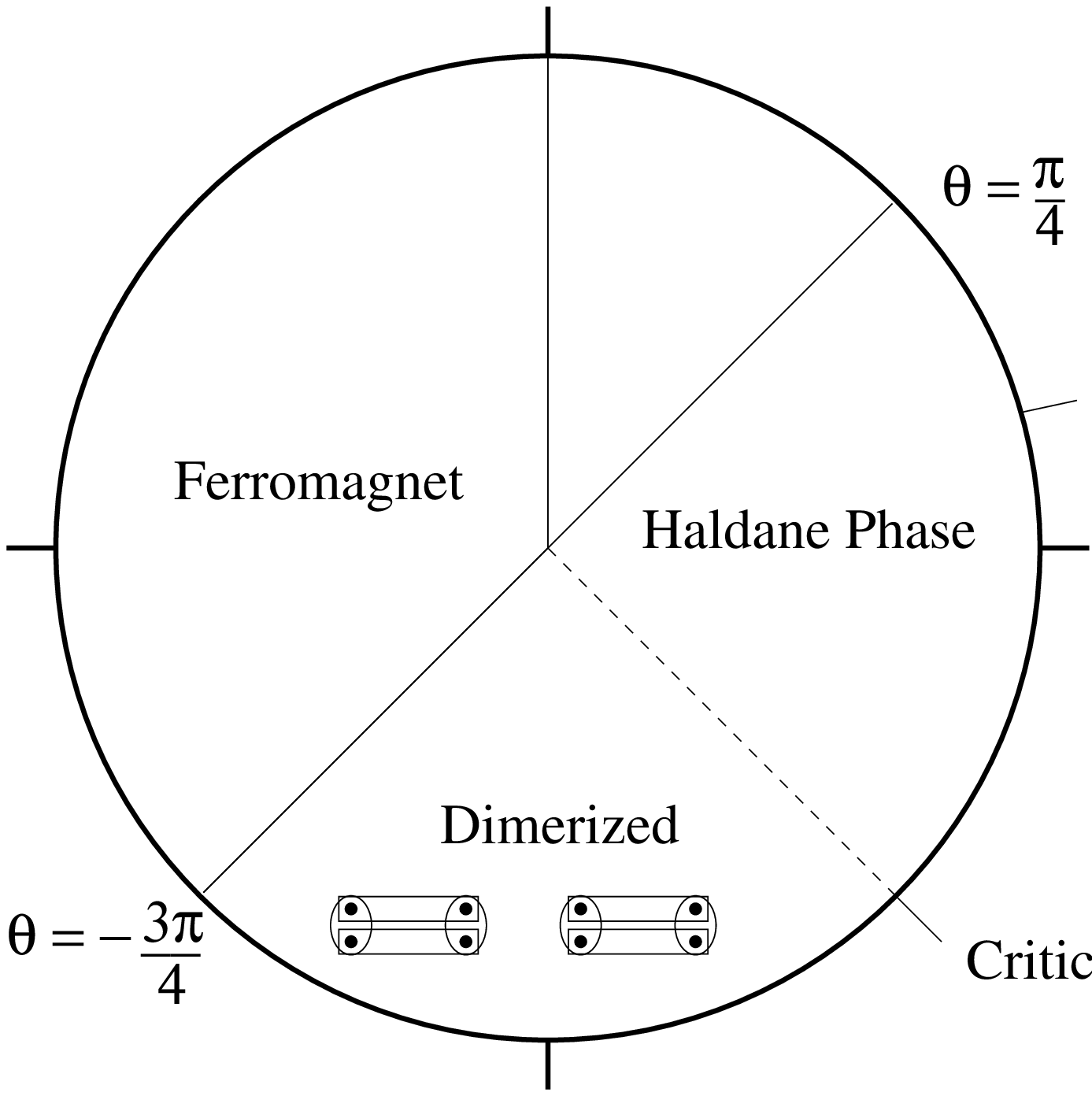}
\caption{Phase diagram for isotropic, nearest-neighbor, S=1 chain, 
parameterized by the angle $\theta$. A 
critical point at $\theta = - \pi/4$ separates the Haldane and dimerized 
phases, both of which are gapped.  At $\theta = \tan^{-1}(1/3)$, the ground 
state is a valence bond solid (VBS), as depicted in the schematic. 
Each oval represents an atom; the two black dots inside the oval are two 
electrons symmetrized into a triplet $S =1$ spin state.  Rectangles 
represent singlet bonds which encompass two electrons on adjacent sites.  
In the dimerized phase, singlet correlations are instead enhanced on 
alternating dimers.}
\label{phasediagram1}
\end{figure}

As a final example we apply the DMRG / finite-size scaling method to the
isotropic spin-$1$ antiferromagnetic chain.  This problem is more challenging
numerically as the on-site Hilbert space now has dimension $D = 3$ instead 
of $D = 2$.  
The most general nearest-neighbor Hamiltonian for the spin-$1$ chain includes
the possibility of a biquadratic spin-spin interaction term:
\begin{equation}
H = \sum_{j=0}^{L-2} \big{[} \cos \theta ~ \vec{S}_j \cdot \vec{S}_{j+1} + 
\sin \theta ~ ( \vec{S}_j \cdot \vec{S}_{j+1} )^2 \big{]}
\label{s1_hamilt}
\end{equation}
The phase diagram can be represented on a circle parameterized by $\theta$ 
as depicted in Fig. \ref{phasediagram1}.
Generically there is a gap to excitations in the antiferromagnetic region 
of the phase diagram, in accord with the Haldane conjecture\cite{Haldane}.
The point $\theta = 0$ corresponds to the usual pure bilinear Heisenberg 
antiferromagnet.  At the point $\tan \theta = 1/3$ the Hamiltonian can
be written as a sum of positive-definite projection operators, and the
exact ground state is the AKLT valence bond solid (VBS)\cite{aklt}. 
Negative $\sin \theta$ favors dimerization, as the energy is minimized by
concentrating singlet correlations on isolated dimers.  The dimerized phase 
also is gapped: a dimer must be broken to generate a spin excitation.  The 
point that separates the dimerized and Haldane phases lies at 
$\theta = - \pi/4$ 
and can be solved exactly by the Bethe ansatz\cite{Takhtajan,Babudjian}. 
The chain is quantum critical at this integrable point.  The ground state is
non-degenerate here as well as in the dimerized and Haldane phases.

\begin{figure}
\epsfxsize=5in \epsfbox{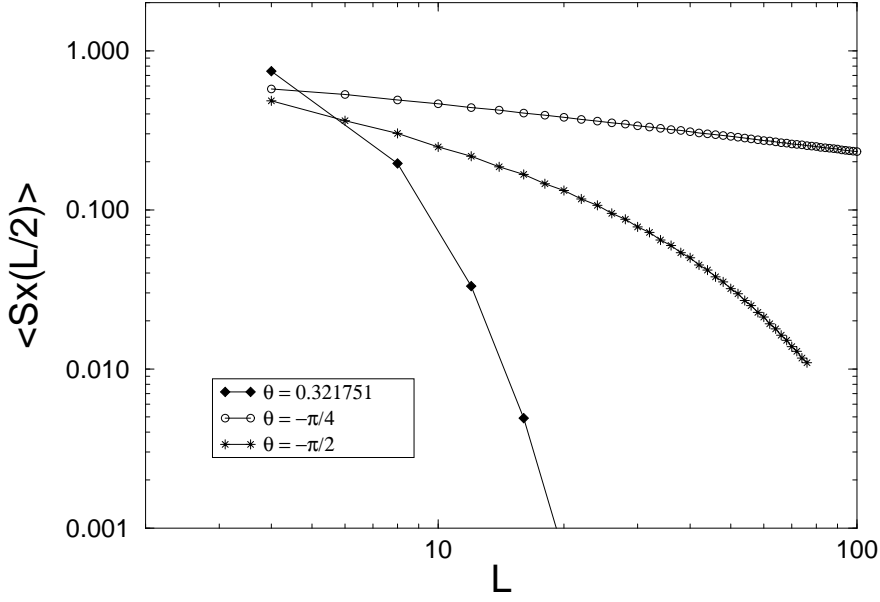}
\caption{Induced spin moment at center of a spin-$1$ chain for different 
values of $\theta$.  Here $h=1$.  Note the power law decay 
at the critical point ($\theta = -\pi/4$).  Exponential decay occurs at 
the AKLT valence bond solid point [$\theta = \arctan(1/3) = 0.321751$] within
the Haldane phase and in the dimerized phase at $\theta = -\pi/2$.  Results 
at the critical point were obtained with $M=256$, while $M=81$ for the 
other two cases.}
\label{s1-sz}
\end{figure}

\begin{figure}
\epsfxsize=5in \epsfbox{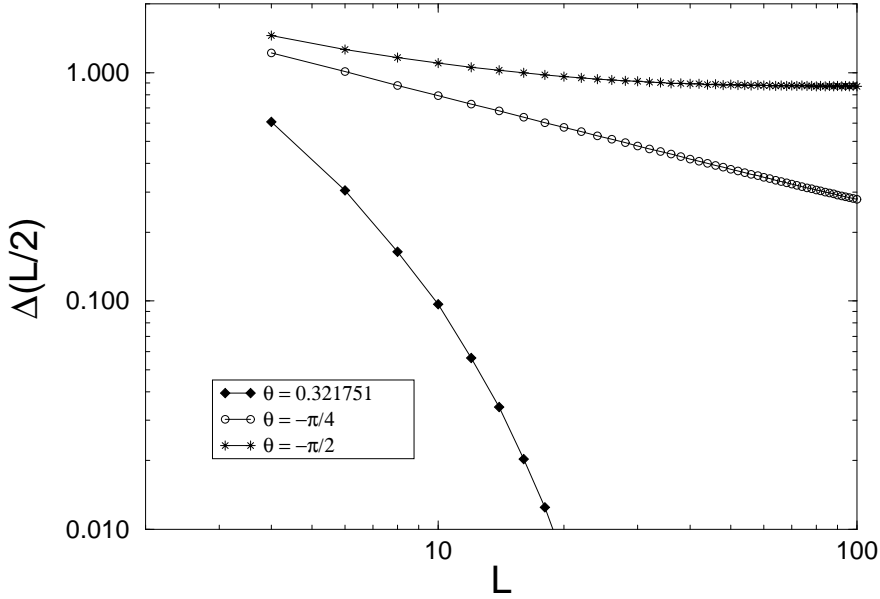}
\caption{Induced dimerization at center of a spin-$1$ chain.  Again we set 
$M=256$ at the critical point and $M=81$ for the two massive phases.  There 
is power law decay at the critical point ($\theta = -\pi/4$) and 
exponential decay at the AKLT point.  As expected the dimerization approaches
a non-zero constant in the dimerized phase ($\theta = -\pi/2$).}
\label{s1-dimer}
\end{figure}

DMRG calculations clearly delineate the two massive phases and the critical 
point separating them, even for relatively small block Hilbert sizes $M$. 
In Figs. \ref{s1-sz} and \ref{s1-dimer}, the blocksize $M=81$ for 
the massive phases. Thus the results are numerically exact up only to chain 
lengths $L=10$.  For chain lengths $L > 10$ the Hilbert space is truncated 
via the DMRG algorithm.  To increase accuracy, results at the critical 
point were obtained with a larger Hilbert size for the blocks, $M=256$.
The induced spin moment at the center of the chain decays exponentially 
in both the Haldane and the dimerized phases, as expected.  The induced 
dimerization at the chain center also decays exponentially in the Haldane 
phase, but approaches a non-zero constant in the dimerized phase as it must.
Power law decay in both observables occurs at the critical point. 
Fitting the $M=256$ data shown in Fig. \ref{s1-dimer} at the critical 
point $\theta = - \pi/4$ we obtain dimerization exponents 
$x_{\Delta} = 0.37 \pm 0.01$ and $y_{\Delta} = 0.3 \pm 0.05$, reflecting 
the apparent presence of a marginal interaction and consequent 
multiplicative logarithmic corrections to scaling.  Likewise, for a field 
of $h = 1.0$ applied to the chain ends, the exponents 
for the spin operator are $x_B = 0.34 \pm 0.01$ and 
$y_B = 0.23 \pm 0.05$.  The values of the exponents compare to the exact 
values\cite{spin1a,spin1b} $x_{\Delta} = 3/8 \approx 0.375$ and 
$x_B = 3/8$.  To the best of our knowledge there are no analytic results at
the integrable point $\theta = - \pi/4$ (which corresponds to a $k=2$ SU(2) 
WZW model) on the size of the logarithmic corrections $y_\Delta$ and $y_B$,
at least for open boundary conditions. 

Finally, we may repeat the analysis of the conformal anomaly described 
above in subsection \ref{subsec:c-half} 
for the case of the spin-1 chain at its critical point.  For $M = 256$ 
and fitting over chain lengths $10 \leq L \leq 26$ we find that the speed
of excitations is $v = 3.69$, $C = 0.508$, and hence $c = 1.05$.  This value
is close to its exact value of 1, demonstrating that the conformal anomaly
can be reliably extracted even from relatively short chains.

\section{Conclusion}
\label{sec:conclusion}
We have presented a simple method for studying critical behavior of quantum 
spin chains. Accurate critical exponents can be extracted. 
For small on-site Hilbert space sizes ($D=2$ for the spin-$1/2$ chain and 
$D=3$ for spin-$1$ chains) the method does not require supercomputers. 
Results can be systematically improved by increasing the size of $M$,
the dimension of the Hilbert space retained in the blocks, up to limits
set by machine memory and speed.  The DMRG method
works best for massive, non-critical, systems, but it is also quite accurate
even at critical points.  Critical exponents can be calculated at percent
level accuracy.  We showed that the leading multiplicative 
logarithmic correction to the scaling of the induced spin moment cancels out 
in the case of the isotropic spin-$1/2$ Heisenberg antiferromagnet.  Thus 
accurate exponents can sometimes be found numerically despite the presence 
of marginal interactions.  

Use of the ``finite-size'' DMRG algorithm might improve the method, but 
good results were obtained with the relatively simpler 
``infinite-size'' DMRG algorithm.  The reason for this is that the 
finite-size scaling method employed here focuses on the scaling of 
observables near the center of the chain only, where the ``infinite-size''
algorithm is particularly accurate.  The method can be 
used to study new systems.  For example, several non-interacting but 
disordered electron systems, like the integer and spin quantum Hall 
transitions, can be described by supersymmetric Hamiltonians\cite{LSM,Tsai1}.  
In a paper which follows\cite{part2}, we employ the combined DMRG/finite-size 
method in combination with analytic calculations to understand the behavior 
of these supersymmetric spin chains. 

\vskip 0.5cm
{\bf Acknowledgments}
We thank I. Affleck, M. P. A. Fisher, V. Gurarie, J. Kondev, M. Kosterlitz,  
A. Ludwig and T. Senthil for useful discussions. 
This work was supported in part by the NSF under 
Grants Nos. DMR-9357613, DMR-9712391.  Computations were
carried out in double-precision C++ on Cray PVP machines at 
the Theoretical Physics Computing Facility at Brown University.

\end{document}